# Partial separability of the Schrödinger equation combined with a Jastrow factor

**Claude Le Sech[1] and Antonio Sarsa[2]**

1) Université Paris-Saclay, CNRS, Institut des Sciences Moléculaires d'Orsay-ISMO (UMR 8214), 91405 Orsay Cedex, France

2) Departamento de Física, Campus de Rabanales Edif. C2, Universidad de Córdoba, E-14071 Córdoba, Spain.

**ABSTRACT:** Describing the Coulomb interactions between electrons in atomic or molecular systems is an important step to help us obtain accurate results for the different observables in the system. One convenient approach is to separate the dynamic electronic correlation, i.e., Coulomb electron-electron repulsion, from the motion of the electrons in the nuclei electric field. The wave function is written as the product of two terms, one accounting for the electron-electron interactions, which is symmetric under identical particle exchange; the other is antisymmetric and represents the dynamics and exchange of electrons within the nuclear electric field. In this work, we present a novel computational scheme based on this idea that leads to an expression of the energy as the sum of two terms. To illustrate the method, we look into few-body Coulombic systems, $H_2$, $H_3^+$ and $Li(1s^2,2s)$, and discuss the possible extension to larger systems. A simple correlation factor, based on the Jastrow exponential term, is employed to represent the dynamics of the electron pairs leading to simple analytical forms and accurate results. We also present and illustrate a different approach with the Li atom based on the partial separability applied to a portion of the atom.

# 1. INTRODUCTION

Accurate wave functions for few-body Coulombic systems can be calculated using different approaches, and this could be considered a closed problem if, for example, large expansions of the wave function in terms of the Slater determinants are used, as in the Configuration Interaction method, see [1], for example. However, producing tailor-made, compact, accurate correlated wave functions is still of great interest, for two main reasons: firstly, it helps us to clarify and understand better the physical meaning of these functions; secondly, they can provide clues for, and help check, some aspects of theoretical studies concerned with sophisticated methods of Quantum Chemistry, such as the exchange-correlation term in Density Functional Theory [2]. Therefore, we should see them as the first step towards generating new tools to treat molecular systems with greater efficiency and accuracy.

One well-known basic problem when studying atoms or molecules is how to describe electron-electron interactions. Different approaches beyond the Hartree-Fock (HF), known as post-HF, such as Coupled Cluster [3,4] or Møller-Plesset [5], have been employed to improve the treatment of electronic correlations.

However, an alternative way to account for electronic correlation was proposed by Hylleraas during the golden age of Quantum Physics. Introducing an explicit dependence on inter-electron distances in the wave function allows for an accurate description of the helium atom. Such explicitly correlated wave functions provide excellent accuracy for the properties of two-electron and three-electron atoms, or molecules, within the framework of a variational optimization, see e.g. [6]. However, when larger systems are considered, the method becomes more complicated and less elegant. Nowadays, explicitly correlated wave functions are widely employed in electronic structure calculations of atoms and molecules by using numerical integration schemes such as those based on the Quantum Monte Carlo methods or other schemes such as the Explicitly Correlated R12/F12 or the Transcorrelated methods the, see e.g. [7-14].

In this work, we represent the dynamic of the electrons in an atom or molecule qualitatively using two modes: electron-nuclei Coulomb interactions – with no interelectron repulsion – and Coulomb repulsion between electrons. Our method is based on both a factorization of the wave function and an analytical manipulation of the expectation values. These have a long tradition of use in variational calculations of different many-fermion systems and computational frameworks, see for example [15]. Thus, the wave function is written as a product of two functions: one factor, which is antisymmetric in particle exchange, accounts for the electron dynamics in the field of the nuclei, while the other, which is symmetric, describes the dynamic correlation among identical particles. Here, we demonstrate that the factorization of the wave function leads to a partial separability of the energy estimator into two integrals depending on the square of the total wave function, using an exponential form or Jastrow for the correlation factor.

To illustrate this method, we focus on few-body Coulombic systems, and present a possible extension to larger systems, and other schemes for partial separability based on considering different components of the total system.

This paper is organized as follows. Section 2 is devoted to the theoretical methods developed in this work. In Section 3 the results here obtained are presented and is devoted to the discussed. The conclusions are provided in Section 4. Atomic units are used throughout this work.

## 2. THEORY

The wave function $\Psi$ describing stationary states of a system of $N$ point-like charges, both nuclei and electrons, can be obtained from the solution of the time independent Schrödinger equation.

$$\left[\sum_{i=1}^{N} -\frac{1}{2m_i} \nabla_i^2 + \sum_{i=1}^{N-1} \sum_{j=i+1}^{N} \frac{q_i q_j}{|\vec{r}_i - \vec{r}_j|}\right] \Psi(\vec{r}_1 \sigma_1, \ldots, \vec{r}_N \sigma_N) = E\, \Psi(\vec{r}_1 \sigma_1, \ldots, \vec{r}_N \sigma_N)$$

where for the $i$-th particle, $\vec{r}_i \sigma_i$ represent the spatial and spin coordinates, $\nabla_i^2$ is the Laplacian operator and $m_i$ and $q_i$ stand for the mass and electric charge in terms of the electron mass and elementary charge respectively.

The basic assumption is that wave function, $\Psi$, can be written as the product of two functions $\Psi = \Phi\Omega$, where no restrictions in the functional forms of these two factors, except for those dictated by Quantum Mechanics, are imposed at this point.

The kinetic energy operator acting on a factorized wave function can be written as follows,

$$\left(\sum_{i=1}^{N} -\frac{1}{2m_i} \nabla_i^2\right)(\Omega\Phi) = \sum_{i=1}^{N} -\frac{1}{2m_i}\left(\Omega\nabla_i^2\Phi + \Phi\nabla_i^2\Omega + 2\vec{\nabla}_i\Omega \cdot \vec{\nabla}_i\Phi\right) \quad (1),$$

and the expectation value of the energy is,

$$E = \int (\Omega\Phi)\left[\sum_{i=1}^{N} -\frac{1}{2m_i}\left(\Omega\nabla_i^2\Phi + \Phi\nabla_i^2\Omega + 2\vec{\nabla}_i\Omega \cdot \vec{\nabla}_i\Phi\right) + V\,\Omega\Phi\right]d\tau, \quad (2),$$

where $V$ is the potential energy operator and $\int d\tau$ indicates the integration over the $N$ particle spatial coordinates and spin and, for purposes of simplicity, we have assumed that the wave function is normalized. If this is not the case, all that is needed is to divide the final expression of the energy by the normalization integral.

By using the following identities

$$\Omega\Phi\,\vec{\nabla}_i\Phi \cdot \vec{\nabla}_i\Omega = \frac{\vec{\nabla}_i\Omega^2 \cdot \vec{\nabla}_i\Phi^2}{4},$$

$$\vec{\nabla}_i\Omega^2 \cdot \vec{\nabla}_i\Phi^2 = \vec{\nabla}_i \cdot \left(\Omega^2\,\vec{\nabla}_i\Phi^2\right) - \Omega^2\,\nabla_i^2\Phi^2,$$

$$\nabla_i^2\Phi^2 = 2\left(\vec{\nabla}_i\Phi \cdot \vec{\nabla}_i\Phi + \Phi\nabla_i^2\Phi\right)$$

the dot product terms in Eq. (2) can be simplified by using the Divergence theorem,

$$\int -\Omega\Phi\,\vec{\nabla}_i\Omega \cdot \vec{\nabla}_i\Phi\,d\tau = \int_{\partial\tau=\sigma} -\frac{(\Omega^2\,\nabla_i^2\Phi^2)}{4} \cdot \hat{n}_i d\sigma_i + \int \frac{\Omega^2(\vec{\nabla}_i\Phi \cdot \vec{\nabla}_i\Phi + \Phi\nabla_i^2\Phi)}{2}d\tau \quad (3)$$

where the surface integrals correspond to the $i$-th particle in the limit $|\vec{r}_i| \to \infty$. These terms vanish, while the terms with the Laplacian of $\Phi$ cancel each other out when substituting Eq. (3) in Eq. (2). Therefore, the energy is the sum of two terms,

$$E = I_1 + I_2, \quad (4)$$

with

$$I_1 = \int (\Omega\Phi)^2 \left[ \sum_{i=1}^{N} -\frac{1}{2m_i} \frac{\nabla_i^2 \Omega}{\Omega} + V_1 \right] d\tau,$$

$$I_2 = \int (\Omega\Phi)^2 \left[ \sum_{i=1}^{N} -\frac{1}{2m_i} \vec{\nabla}_i \ln|\Phi| \cdot \vec{\nabla}_i \ln|\Phi| + V_2 \right] d\tau \quad (5)$$

where we have split the potential into two contributions,

$$V = V_1 + V_2.$$

Equations (4) and (5) illustrate the partial separation between the functions $\Omega$ and $\Phi$, as the coupling term, $\vec{\nabla}_i \Omega \cdot \vec{\nabla}_i \Phi$ of Eq. (2), does not appear in the final form of the expectation value of the energy. The first contribution to the energy, $I_1$, represents the kinetic energy of the motion associated to function $\Omega$, whereas the second term, $I_2$, includes the log gradient of the $\Phi$ function. The separation of the potential energy is arbitrary, and any contribution can be cast either in $I_1$ or $I_2$ at will. No further hypotheses have been made for either $\Omega$ or $\Phi$, so that any choice based on physical grounds relative to the nature or role of the particles or on computational convenience can be considered.

In this work, we set $\Omega$ to include correlations between identical particles or with similar dynamics. For example, for a molecule can be taken as $\Omega$ as a correlation factor depending on electron-electron and nucleus-nucleus distances symmetric under the exchange of identical particles written in a convenient exponential form,

$$\Omega(\vec{r}_1, \ldots, \vec{r}_N) = \exp\left[\sum_{i<j}^{N_e} u(r_{ij})\right] \exp\left[\sum_{i_A=1}^{N_A} \sum_{j_B=1}^{N_B} w_{AB}(r_{i_A j_B})\right] \exp\left[\sum_{i_A=1}^{N_A} \sum_{j_C=1}^{N_C} w_{AC}(r_{i_A j_C})\right] \ldots,$$

with $N_e$ the number of electrons, $N_A$, the number A-type of nuclei, $N_B$, B-type of nuclei etc.; $r_{ij}$ the distance between the, $i$, and, $j$, particles; $u$, and $w_{XY}$, different functions, with the product of exponential functions extended to all different nuclear pairs.

For the $\Phi$ factor a simple choice is to use spin orbitals depending on electron-nuclear distances and with the proper symmetry under the exchange of identical particles,

$$\Phi(\vec{r}_1\sigma_1, \ldots, \vec{r}_N\sigma_N) = \mathcal{A}\left[\varphi_{\alpha_1}(\vec{r}_1\sigma_1) \ldots \varphi_{\alpha_N}(\vec{r}_N\sigma_N)\right]$$

with $\mathcal{A}$ the symmetrization and antysimmetrization operator for respectively identical bosonic and fermionic particles [16]. Other correlation mechanisms can be included in this term without increasing the complexity of the problem as we shall illustrate below.

One of the interests of the present method to describe the electronic correlation accurately relies on the possibility of obtaining the energy with a modest effort starting from Eq. (5). For an $N$-electron atom in the clamped nucleus approximation, the use of a correlation factor calculated as the product of the electron pair functions,

$$\Omega(\vec{r}_1, \ldots, \vec{r}_N) = e^{\sum_{i<j}^{N} u(r_{ij})},$$

and $V_1$ as the electronic repulsion potential, leads to a simple expression of $I_1$,

$$I_1 = \frac{1}{2} \sum_{i=1}^{N} \int (\Omega\Phi)^2 \left[ -\sum_{i \neq j}^{N} \nabla_i^2 u(r_{ij}) - \left( \sum_{i \neq j}^{N} \vec{\nabla}_i u(r_{ij}) \right)^2 + \sum_{j \neq i}^{N} \frac{1}{r_{ij}} \right] d\tau$$

The function $\Phi$ describes the motion of the electrons in the electric field of the nuclei and $V_2$ is the electron-nucleus interaction, so that $I_2$ in Eq. (5) is,

$$I_2 = \sum_{i=1}^{N} \int (\Omega\Phi)^2 \left[ \frac{1}{2} \left( \vec{\nabla}_i \ln|\Phi| \right)^2 - \frac{Z}{r_i} \right] d\tau$$

with $Z$ being the nuclear charge and $\Phi$ an antisymmetric mean field or Configuration Interaction wave function depending on particle position and spin that can be written as a linear combination of Slater determinants so that and the log derivative with respect to each particle coordinates, can be computed efficiently using the matrix of cofactors. Note that only first derivatives of $\Phi$ are required.

It is known that using a three body Jastrow factor that correlates two electrons and a nucleus, and an even elaborated general correlation factor (three electrons and a nucleus) leads to lower energies, see e.g. [17] and references therein. These factors lead to complex forms of the $I_1$ integral that require calculations involving numerical integrations a larger number of dimensions than those needed when using the two-body correlation factor. Recently, in the context of the transcorrelated approach a new three body, electron-electron-nucleus, correlation factor with the electron-electron and electron-nucleus terms factorized has been proposed [18] with orbital optimization in presence of correlations. A similar idea is used here including the electron-electron are terms in the $\Omega$ factor and the electron-nucleus ones in $\Phi$, with all the functions optimized simultaneously as we show below. As a difference, the electron-nucleus-electron term of this work is intended to account for the screening of the nuclear field acting on one electron due to the presence of the other electron while in [18] the three-body correlation factor is designed to vanish around each nucleus in order to include in the Jastrow factor electron-electron correlations in the valence region. Within the scheme of this work, both kind of three-body correlations can be included with no extra computational demand.

Finally, another possibility of partial separability that will be explored here is, starting from physical considerations relative to the nature or role of particles, to make a separation of coordinates. This option will be illustrated here in the ground state of the Li atom.

The calculation of the energy value *E* necessitates a multidimensional integration. In the present work, a non-statistical method based on Hammersley points [19] is used. For systems with more electrons, a better choice is to consider using the Quantum Variational Monte Carlo approach. All the integrals are weighted by the square of the wave function, as required, to obtain a probability distribution function.

To illustrate the present approach, we look first at well-known few-electron systems such as the $H_2$ molecule without the Born-Oppenheimer approximation, the $H_3^+$ molecular ion in two different nuclear geometries and the Li atom. A symmetric Jastrow term for particles exchange is chosen for $\Omega$, while $\Phi$ is antisymmetric under electron exchange and is written in terms of screened orbitals. We also discuss other possible partial separations and present results applied to the lithium atom.

## 3. RESULTS

*$H_2$ molecule*

With the present scheme, the $H_2$ molecule is studied without the Born-Oppenheimer approximation. The time independent Schrödinger equation can be written in relative Jacobi coordinates as [20]

$$\left[-\frac{1}{2\varepsilon}\nabla_{r_1}^2 - \frac{1}{2\varepsilon}\nabla_{r_2}^2 - \frac{1}{2\mu_R}\nabla_R^2 + V\right]\Psi(\vec{r}_1, \vec{r}_2, \vec{R}) = E\Psi(\vec{r}_1, \vec{r}_2, \vec{R})$$

where the mass polarization term has not been included. The potential energy operator, *V*, is

$$V = -\frac{1}{r_{1A}} - \frac{1}{r_{1B}} - \frac{1}{r_{2A}} - \frac{1}{r_{2B}} + \frac{1}{R} + \frac{1}{r_{12}}$$

with

$$r_{iA} = |\vec{r}_i - \vec{R}_A|, \quad r_{iB} = |\vec{r}_i - \vec{R}_B|; \quad i = 1, 2$$

and

$$R = |\vec{R}_A - \vec{R}_B|, \quad r_{12} = |\vec{r}_1 - \vec{r}_2|$$

The origin is at the center of mass of the nuclei, $\vec{r}_1$ and $\vec{r}_2$ stand for the electronic coordinates, with $\vec{R}_A$ and $\vec{R}_B$ the coordinates of the nuclei situated on the z-axis. Here, we use the value $\varepsilon \approx 1$ $m_e$ and $\mu_R = 918.075$ $m_e$.

The factorization considered here for the wave function is to include in $\Omega$ electron-electron and nucleus-nucleus correlations. A Jastrow or exponential form has been employed here for the explicit dependence on both the interelectronic $r_{12}$ and the internuclear distance $R$

$$\Omega(r_{12}, R) = e^{u(r_{12}) + w(R)}$$

We employ the simple function proposed by Boys and Handy [21] for the interelectronic factor that provides the correct behavior at large electronic distances and with a proper choice of the parameter can satisfy the electron-electron cusp so that the local energy does not diverge at the electron coalescence point,

$$u(r_{12}) = \frac{c \, r_{12}}{1 + b \, r_{12}}$$

while for the internuclear factor

$$w(R) = -\delta \, (R - R_0)^2$$

is used to describe for the lowest vibrational state $v = 0$ of the nuclei. The parameters $c$, $b$, $\delta$ and $R_0$ are to be fixed variationally.

For the $\Phi$ factor, we have used a description in terms of screened atomic electronic orbitals, as that of Refs. [20, 22],

$$\Phi(\vec{r}_1, \vec{r}_2, \vec{R}) = \varphi(\vec{r}_1, \vec{R}) \, \varphi(\vec{r}_2, \vec{R})$$

with $\varphi(\vec{r}_i, \vec{R})$ a three-body wave function, the two nuclei and one electron given by

$$\varphi(\vec{r}_i, \vec{R}) = e^{-(\alpha + \beta R)\lambda_i} \cosh\{[\alpha - (1 - \beta)R]\mu_i\}, \quad i = 1, 2$$

and where $\lambda_i$ and $\mu_i$ are the confocal elliptic coordinates of the i-th electron.

$$\lambda_i = \frac{r_{iA} + r_{iB}}{R}, 1 \leq \lambda_i \leq \infty; \quad \mu_i = \frac{r_{iA} - r_{iB}}{R}, -1 \leq \mu_i \leq 1$$

and $\alpha$ and $\beta$ are variational parameters. This non-adiabatic wave function describes the electronic motion for any nuclear distance providing accurate results for the molecule when used along with the other factor under different environments [20]. The functional form selected for the orbital includes and electron-nucleus-nucleus correlations in the form of an effective nuclear that depends on the nuclear separation, that changes due to the vibrational motion of the two nuclei.

The potential energy separation is done for the same interparticle distance as that of the wave function.

$$V_1 = \frac{1}{R} + \frac{1}{r_{12}}$$

With this choice, we obtain,

$$I_1 = \int (\Omega\Phi)^2 \left[ \left(-\tfrac{1}{2}\nabla_{\vec{r}_1}^2 - \tfrac{1}{2}\nabla_{\vec{r}_2}^2\right) u(r_{12}) - \tfrac{1}{2\mu_R} \nabla_{\vec{R}}^2 w(R) + V_1 \right] d\tau$$

$$= \frac{\delta}{\mu_R} + \int (\Omega\Phi)^2 \left\{ \left[ -\frac{2\delta^2(R-R_0)^2}{\mu_R} + \frac{1}{R} \right] + \left[ \frac{-c}{(1+br_{12})^3} \left( \frac{2}{r_{12}} + \frac{c}{1+br_{12}} \right) + \frac{1}{r_{12}} \right] \right\} d\tau \quad (6)$$

and

$$I_2 = \int (\Omega\Phi)^2 \left[ \sum_{i=1}^{2} \tfrac{1}{2} \left(\vec{\nabla}_{\vec{r}_i} \ln|\Phi|\right)^2 + \frac{1}{2\mu_R} \left(\vec{\nabla}_{\vec{R}} \ln|\Phi|\right)^2 - \frac{1}{r_{1A}} - \frac{1}{r_{1B}} - \frac{1}{r_{2A}} - \frac{1}{r_{2B}} \right] d\tau$$

The integration is performed in the coordinates $\vec{r}_1, \vec{r}_2$ and $\vec{R}$.

It must be noted that for small $r_{12}$ values, the term within the second bracket is ≈ $-2c/r_{12} + 1/r_{12}$, so that choosing $c = 0.5$ $a_0^{-1}$ cancels out the regular singularities of the Coulomb field for antiparallel-spin electrons as those of this molecule. This value will be used throughout this work. The other parameters in the wave function, $R_0$, $b$, $\delta$, $\alpha$ and $\beta$, are fixed variationally obtaining an energy of $E(H_2) = -1.1602$ $E_h$ with $R_0 = 1.4$ $a_0$, $b = 0.25$ $a_0^{-1}$, $\delta = 9.1$ $a_0^{-2}$, $\alpha = 0.16$ and $\beta = 0.58$ $a_0^{-1}$. The dissociation energy from the lowest vibrational state v = 0 is $D_0 = 0.1602$ $E_h$, to be compared to the accurate value $D_0' = 0.1639$ $E_h$ [23]. The expectation value of the nuclear distance here obtained is $<R> \approx 1.4$ $a_0$.

The Born-Oppenheimer approximation can be retrieved with our model by setting $\mu_R \to \infty$ obtaining an energy of $E(\mu_R \to \infty) = -1.1702$ $E_h$. The dissociation energy from the bottom is $D_e = 0.1702$ $E_h$, close to the accurate value $D_e' = 0.1744$ $E_h$ [23].

The ratio $\delta/\mu_R = 9.1/918.075 \approx 0.0099$ $E_h$ is very close to $D_e - D_0 = 0.01$ $E_h$, and $D_e' - D_0' = 0.0105$ $E_h$. This agreement supports the idea that $\delta/\mu_R$ represents most of the stretching energy. The corresponding vibrational frequency is $\omega = 4390$ cm$^{-1}$.

*Molecular ion $H_3^+$*

The $H_3^+$ molecular ion is presented here in the Born-Oppenheimer approximation, using both equilateral and linear geometry. The origin of the coordinate system is located at the center of mass of the nuclei. For a fixed distance $d$ between the nuclei A, B & C, the potential energy is

$$V_d^k(\vec{r}_1, \vec{r}_2) = \frac{k}{d} + \frac{1}{r_{12}} + \sum_{i=1}^{2} -\frac{1}{r_{iA}} - \frac{1}{r_{iB}} - \frac{1}{r_{iC}},$$

with $k = 3$ and $k = 5/2$ for equilateral and linear geometry, respectively.

The electronic Schrödinger equation for a given nuclear distance $d$ can be written in either geometry as

$$\left[-\frac{1}{2}\nabla_{r_1}^2 - \frac{1}{2}\nabla_{r_2}^2 + V_d^k\right] \Psi_d^k(\vec{r}_1, \vec{r}_2) = E_d^k \, \Psi_d^k(\vec{r}_1, \vec{r}_2).$$

As before, we have used the Boys and Handy Jastrow form, depending on the interelectronic distance, $r_{12}$, for $\Omega$,

$$\Omega(r_{12}) = e^{\frac{cr_{12}}{1+br_{12}}}$$

with $c$ and $b$ variational parameters. The same functional form is employed in both cases.

The $\Phi$ factor describes the motion of the electrons in the electric field of the nuclei, depends on the geometry of the molecule, and it is written here in terms of screened orbitals,

$$\Phi_d^k(\vec{r}_1, \vec{r}_2) = \varphi_d^k(\vec{r}_1) \, \varphi_d^k(\vec{r}_2).$$

In the triangular geometry, the following form for the orbitals is used

$$\varphi_d^t(\vec{r}_i) = e^{-Z_{eff} \, r_{iA} - e \, r_{iB} - e \, r_{iC}} + e^{-Z_{eff} \, r_{iC} - e \, r_{iA} - e \, r_{iB}} + e^{-Z_{eff} \, r_{iB} - e \, r_{iC} - e \, r_{iA}}, \qquad i = 1, 2$$

where $Z_{eff}$ and $e$ are variational parameters.

A more general orbital, including a third variational parameter, is employed for the linear geometry,

$$\varphi_d^l(\vec{r}_i) = e^{-Z_{eff} \, r_{iA} - e \, r_{iB} - e' \, r_{iC}} + e^{-Z_{eff} \, r_{iC} - e \, r_{iA} - e' \, r_{iB}} + e^{-Z_{eff} \, r_{iB} - e \, r_{iC} - e' \, r_{iA}}, \qquad i = 1, 2$$

The minimum energy for the equilateral geometry is $E = -1.3427$ $E_h$, which corresponds to an equilibrium nuclear distance of $d = 1.63$ $a_0$, obtained with $b = 0.23$ $a_0^{-1}$, $c = 0.5$ $a_0^{-1}$, $Z_{eff} = 1.28$ $a_0^{-1}$ and $e = 0.21$ $a_0^{-1}$. This result is in agreement with the accurate values reported in Ref. [24] $E = -1.3427$ $E_h$, $d = 1.63$ $a_0$, calculated using a wave function with ten variational parameters.

In the case of a linear molecule, the optimized values of the parameters are $b = 0.21$ $a_0^{-1}$, $c = 0.5$ $a_0^{-1}$, $Z_{eff} = 1.27$ $a_0^{-1}$, $e = 0.17$ $a_0^{-1}$ and $e' = 0.24$ $a_0^{-1}$; giving an equilibrium distance and energy of $d = 1.53$ $a_0$ and $E = -1.2776$ $E_h$, in close agreement with the result reported by Conroy [25] $d = 1.54$ $a_0$ and $E = -1.278$ $E_h$.

To illustrate the interest of the partial separability in the study of molecular structure, we report the $I_1$ and $I_2$ values for the $H_3^+$ molecular ion in both equilateral and linear geometries and the $H_2$ molecule. The expectation values of the electron-nucleus potential energy, $<V_{en}>$, and the interelectronic separation, $<r_{12}>$, are also shown.

Table 1. Numerical values of $I_1$, $I_2$, $<V_{en}>$ and $<r_{12}>$ for the $H_3^+$ molecular ion in equilateral and linear geometries and $H_2$ molecule.

|  | $I_1$ | $I_2$ | $<V_{en}>$ | $<r_{12}>$ |
|---|---|---|---|---|
| $H_3^+$ equilateral | 0.2924 $E_h$ | -3.4747 $E_h$ | -5.17 $E_h$ | 1.96 $a_0$ |
| $H_3^+$ linear | 0.2469 $E_h$ | -3.1374 $E_h$ | -4.72 $E_h$ | 2.23 $a_0$ |
| $H_2$ | 0.3028 $E_h$ | -2.1837 $E_h$ | -3.67 $E_h$ | 2.14 $a_0$ |

For minimum energy geometry of the $H_3^+$ molecular ion, the equilateral configuration, the $I_1$ energy is higher than for the linear molecule, whereas $I_2$ is lower for the triangle compared to the linear. Recall that the electron-electron energy is included in $I_1$ while the electron-nuclei energy in $I_2$. The $<r_{12}>$ values show that in the triangular configuration the electrons are, on average, closer together than in the linear one. This makes the value of $I_1$ higher in the former because of both the larger electrostatic repulsion and the greater contribution to the total kinetic energy due to the higher curvature of this factor of the wave function. However, these effects are counterbalanced in the total energy by the contribution of $I_2$, containing both the nuclear attraction and the kinetic energy of the mean-field part of the wave function, which makes the energy lower when the nuclei have triangular geometry. For the sake of comparison, we also report the values obtained for the H$_2$ molecule that presents a larger value of $I_1$ than in the case of than in the linear geometry of H$_3^+$ because of the larger interelectron distance in the later. The $I_2$ energy is much lower in H$_3^+$ as expected due to the presence of a third nucleus.

*The lithium atom*

The lithium atom is considered in its ground state Li(1s$^2$2s). The Boys and Handy correlation function,

$$u(r_{ij}) = \frac{c\, r_{ij}}{1 + b\, r_{ij}}, \quad i,j = 1,2,3$$

is employed in Ω. The Φ factor is written as follows,

$$\Phi(r_1, r_2, r_3) = \varphi_{1s}(r_1)\,[\varphi_{1s}(r_2)\varphi_{2s}(r_3) - \varphi_{2s}(r_2)\varphi_{1s}(r_3)]\,\cosh(\lambda r_1)\cosh(\lambda r_2)\cosh(\lambda r_3),$$

where the $\varphi_{ns}(r)$ functions are hydrogenic orbitals with the nuclear charge $Z = 3$. This wave function produces an identical expectation value of any spin-independent operator as the full antisymmetric wave function [26, 27]. The factor with the hyperbolic functions describes the screening of the nuclear charge, with $\lambda$ the screening coefficient, as well as being consistent with the cusp condition and providing accurate results for few-electron atoms [28].

The wave function has three parameters, with $\lambda$ and $b$ fixed variationally and we take the value $c = 0.5$ a$_0^{-1}$ to fulfill Kato's cusp conditions for two electrons with different spin. The minimum value for the energy obtained here is $E = -7.4710$ $E_h$, which is compared to the Hartree-Fock result $E = -7.4327$ $E_h$ and the accurate result E $= -7.4780$ $E_h$ [29]. The parameter values are $\lambda = 0.67$ a$_0^{-1}$, $b = 0.82$ a$_0^{-1}$. This wave function with a symmetric Jastrow factor does not present spin contamination but only the antiparallel-spin electron cusp is fulfilled.

With the choice of this simple effective two-parameters wave function the energy value is 0.007 $E_h$ above the estimated exact value. More elaborated Jastrow factors for atoms, molecules, and solids, are proposed in the literature for example [18, 30]. The aim of the present work is to show the interest of the partial separability of the Schrödinger equation using simple ansatz, and the possibility to consider sophisticated Jastrow factor with more parameters when a better accuracy is required.

In Table 2 we summarize the results of this work. More accurate results can be obtained with the present scheme by using more complex expressions for either Ω or Φ, for example, by considering more general parameterizations of the correlation factor or by including more elaborated three-body correlations (e-e-n), expanding the antisymmetric part to more Slater determinants, or using other model functions based on correlated electron pairs or geminals. It is worth noting that compact and accurate wave functions such as those obtained in this work are especially suited to guiding the simulation in Quantum Monte Carlo calculations such as Diffusion or Reptation Quantum Monte Carlo methods.

**Table 2. Summary of the different results reported in this work.**

|  | $Z_{eff}$ | c | e, e' | b | d | Energy |
|---|---|---|---|---|---|---|
| H-H (BO) | $\alpha + \beta d$ | 0.5 | 0.17 | 0.22 | 1.4 | -1.1702 |
|  |  |  |  |  |  | -1.1744[1] |
| H-H (non BO) | $\alpha + \beta R$ | 0.5 | 0.17 | 0.25 |  | -1.1602 |
|  |  |  |  |  |  | $\omega$ = 4390cm$^{-1}$ |
| H-H-H$^+$ (equilateral) | 1.28 | 0.5 | 0.21 | 0.23 | 1.63 | -1.3427 |
|  |  |  |  |  |  | -1.3432[2] |
| H-H-H$^+$ (linear) | 1.28 | 0.5 | 0.17 | 0.21 | 1.53 | -1.2750 |
|  |  |  | 0.24 |  |  | -1.278[3] |
| Li(1s$^2$2s) | 3 | 0.5 | $\lambda$=0.67 | 0.82 |  | -7.4710 |
|  |  |  |  |  |  | -7.4780[4] |

[1]Ref [23]; [2]Ref [24]; [3]Ref [25]; [4]Ref [29]

***Different choice of functions $\Phi$, $\Omega$ to study the lithium atom.***

One of the interests of the present method lies in the flexibility to define the factors $\Phi$ and $\Omega$ to carry out the partial separability of the energy. This opens up the possibility of describing quantitatively a sub-part of an atomic or molecular system by including the subsystem of interest in the $\Omega$ factor.

We illustrate this here by studying the electron pair with the same spin in the Li atom, namely the (2,3) pair, with the choice of coordinates of this work. The variational wave function employed here gives us

$$\Omega(\vec{r}_2, \vec{r}_3) = \phi(r_2, r_3)\, e^{\frac{cr_{23}}{1+br_{23}}}$$

with

$$\phi(r_2, r_3) = [\varphi_{1s}(r_2)\varphi_{2s}(r_3) - \varphi_{2s}(r_2)\varphi_{1s}(r_3)].$$

In line with this, the potential to be included in $I_1$ is

$$V_1(\vec{r}_2,\vec{r}_3) = -\frac{Z}{r_2} - \frac{Z}{r_3} + \frac{1}{r_{23}}$$

Therefore, $I_1$ represents now the energy of the electron pair (2,3), which are electrons with the same spin and different spatial orbitals, in the electric field of the nucleus with charge $Z$ and their mutual electrostatic repulsion.

Next, the $\Phi$ function is,

$$\Phi(\vec{r}_1,\vec{r}_2,\vec{r}_3) = \varphi_{1s}(r_1)\cosh(\lambda r_1)\cosh(\lambda r_2)\cosh(\lambda r_3)e^{\frac{cr_{12}}{1+br_{12}}+\frac{cr_{31}}{1+br_{31}}}$$

and $V_2$

$$V_2(\vec{r}_1,\vec{r}_2,\vec{r}_3) = -\frac{Z}{r_1} + \frac{1}{r_{12}} + \frac{1}{r_{31}}$$

Thus, $I_2$ is the energy of one electron in the *1s* orbital in the nuclear field and the average electrostatic field generated by the electronic cloud of the other electrons in the atom.

As the $\varphi_{ns}(r)$ functions in the $\Omega$ factor are hydrogenic orbitals and the integrand in $I_1$ is symmetric in the coordinates of electrons 2 and 3, the integral can be simplified as follows,

$$I_1 = \varepsilon_{1s} + \varepsilon_{2s} + I_{1R},$$

with $\varepsilon_{ns} = -Z^2/(2n^2)$ the hydrogenic single particle energies, and

$$I_{1R} = \int (\Omega\Phi)^2 \left[ -\nabla_2^2 u(r_{23}) - (\vec{\nabla}_2 u(r_{23}))^2 - 2\frac{\vec{\nabla}_2 \phi(r_2,r_3)}{\phi(r_2,r_3)} \cdot \vec{\nabla}_2 u(r_{23}) + \frac{1}{r_{23}} \right] d\tau$$

Thus, $I_1$ is the sum of the single particle energies of the occupied orbitals, plus a correction in the non-interacting electron model provided by the integral term above. The optimal parameters of the wave function of this work produce a numerical value of the integral $I_{1R} = 0.3560\ E_h$, giving $I_1 = -5.2690\ E_h$.

The $I_2$ energy can be split into two contributions, one arising from the kinetic energy, $I_{2T}$, and the other from the potential energy, $I_{2V}$.

$$I_{2T} = \frac{1}{2}\int (\Omega\Phi)^2 \left[ (\vec{\nabla}_1 \ln|\Phi|)^2 + 2(\vec{\nabla}_2 \ln|\Phi|)^2 \right] d\tau,$$

$$I_{2V} = \sum_{i=1}^{N}\int (\Omega\Phi)^2 \left[ -\frac{Z}{r_1} + 2\frac{1}{r_{12}} \right] d\tau$$

With the wave function of this work, we have obtained $I_{2T} = 3.8384\ E_h$ and $I_{2V} = -6.0324\ E_h$, giving $I_2 = I_{2T} + I_{2V} = -2.2020\ E_h$.

With these results, we can recover the value of the total energy $E = I_1 + I_2 = -7.4710\ E_h$.

## 4. CONCLUSIONS

Partial separability is a general property of factorized wave functions used to obtain approximate solutions of the Schrödinger equation. It provides the possibility to treat dynamic correlations and the motion of the electrons in the electric field of the nuclei in different ways. Total wave function is written as a product of two functions: one accounting for the electron pairs, where only the electron-electron repulsion is considered, and the other as independent electrons in a mean field, including the fermionic symmetry. Following this approach, the treatment of the electron pair correlations is similar regardless of whether atoms or molecules are considered, i.e., they are not dependent on the system. The specificity of the system of interest lies in another factor describing the electron dynamics in the Coulomb field of the nuclei. The mathematical task is limited to derivatives of terms relative to only one or two electron coordinates. This approach involves the calculation of multivariate integrals containing the square of the total wave function. The method has been illustrated for the $H_2$ molecule, $H_3^+$ molecular ion and the Li atom, providing accurate results.

The Quantum Monte Carlo method should be relevant and efficient when much more particles are present, in order to determine accurate values. Partial separability of the wave function with the choice of the factors proposed here, based on physical grounds, leads to a form of the energy that is especially suited for use in a Variational Monte Carlo calculation, and which provides accurate results for the energy.

Different schemes of partial separation of the wave function can be developed within the formal framework presented here in such a way that properties of part of the whole system can be extracted. This could be useful in different problems in Molecular Physics or Chemistry, for instance, to analyze the dynamics of some special atoms in the molecule. The determination of the $H_2$ stretching frequency presented above illustrates the partial separability of the nuclei motion from the molecular electron cloud. We have illustrated this application of partial separability by isolating the motion of a pair of equal spin electrons in the Li atom.

**ACKNOWLEDGMENTS**

The work of one of the authors (AS) was partially supported under Grant PID2020-114807GB-I00 by the Spanish MCIN/ AEI /10.13039/501100011033.